\documentclass[multphys,vecphys]{svmult}

\usepackage{graphicx}
\usepackage{makeidx}
\usepackage{multicol}
\usepackage{footmisc}
\usepackage{amssymb}
\usepackage{amsmath}
\usepackage{epsfig}
\usepackage{rotating}

\makeindex
\begin{document}

\title*{%
The Canonical Approach to Quantum Gravity:\\
General Ideas and Geometrodynamics\protect\footnote{%
To appear in: Erhard Seiler \& Ion-Olimpiu Stamatescu (editors):
`Approaches To Fundamental Physics -- An Assessment Of 
Current Theoretical Ideas' (Springer Verlag, to be published).}
\index{Geometrodynamics}}
\titlerunning{%
Canonical Quantum Gravity}
\author{%
Domenico Giulini
\inst{1} 
and Claus Kiefer
\inst{2}
}
\institute{%
Physikalisches Institut, Universit\"at Freiburg\\ 
Hermann-Herder-Stra{\ss}e\,3, D-79104 Freiburg, Germany
\and
Institut f\"ur Theoretische Physik, Universit\"{a}t zu K\"oln\\ 
Z\"ulpicher Stra{\ss}e\,77, D-50937 K\"oln, Germany
}

\maketitle

\begin{abstract}
We give an introduction to the canonical formalism of Einstein's 
theory of general relativity. This then serves as the starting 
point for one approach to quantum gravity called quantum 
geometrodynamics. The main features and applications of this
approach are briefly summarized.
\end{abstract}

\section{Introduction}

The really novel feature of General Relativity\index{General Relativity} 
(henceforth abbreviated GR), as compared to other field theories in physics, 
is that spacetime is not a fixed background arena that merely stages 
physical processes. Rather, spacetime is itself a dynamical entity, 
meaning that its properties depend in parts on its specific matter 
content. Hence, contrary to the Newtonian picture, in which spacetime 
acts (via its inertial structure) but is not acted upon by matter, 
the interaction between matter and spacetime now goes both ways. 

Saying that the spacetime is `dynamic' does not mean that it 
`changes' with respect to any given external time. Time is clearly 
within, not external to spacetime. Accordingly, solutions to Einstein's 
equations, which are whole spacetimes, do not as such describe
anything evolving. In order to take such an evolutionary form,
which is, for example, necessary to formulate an initial value problem,
\index{initial value problem}   
we have to re-introduce a notion of `time' with reference to which 
we may speak of `evolution'. This is done by introducing a structure 
that somehow allows to split spacetime into space and time.

\begin{figure}[htb]
\centering\epsfig{figure=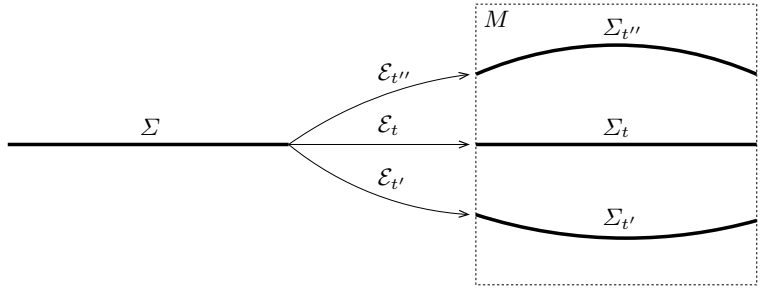,width=0.86\linewidth}
\put(-235,57){\small $\Sigma$}
\put(-145,58){\small $\mathcal{E}_t$}
\put(-145,38){\small $\mathcal{E}_{t'}$}
\put(-145,78){\small $\mathcal{E}_{t''}$}
\put(-60,57){\small $\Sigma_t$}
\put(-60,23){\small $\Sigma_{t'}$}
\put(-60,95){\small $\Sigma_{t''}$}
\put(-105,98){\small $M$}
\caption{\small Foliation of spacetime $M$ by a one-parameter 
family of embeddings $\mathcal{E}_t$ of the 3-manifold $\Sigma$ 
into $M$. $\Sigma_t$ is the image in $M$ of $\Sigma$ under 
$\mathcal{E}_t$. Here the leaf $\Sigma_{t'}$ is drawn to lie to 
the past and $\Sigma_{t''}$ to the future of $\Sigma_t$. 
\label{fig:Nico1}}
\end{figure}

Let us explain this in more detail: suppose we are give a spacetime,
that is, a four dimensional differentiable manifold $M$ with Lorentzian 
metric $g$. We assume that $M$ can be foliated by a family 
$\{\Sigma_t\mid t\in \mathbb{R}\}$ of spacelike leaves.  That is, 
for each number $t$ there is an embedding of a fixed 3-dimensional 
manifold $\Sigma$ into $M$, 
\begin{equation}
\label{eq:SigmaEmbedding}
\mathcal{E}_t:\Sigma\rightarrow M\,,
\end{equation}
whose image $\mathcal{E}_t(\Sigma)\subset M$ is just $\Sigma_t$,
which is a spacelike submanifold of $M$; see Fig.\,\ref{fig:Nico1}. 
It receives a Riemannian metric by restricting the Lorentzian metric 
$g$ of $M$ to the tangent vectors of $\Sigma_t$. This can be expressed 
in terms of the 3-manifold $\Sigma$. If we endow $\Sigma$ with the 
Riemannian metric 
\begin{equation}
\label{eq:SpatialMetric}
h_t:=\mathcal{E}_t^*g\,,
\end{equation}
then $(\Sigma,h_t)$ is isometric to the submanifold $\Sigma_t$ 
with the induced metric. 

Each three dimensional leaf $\Sigma_t$ now corresponds to an 
instant of time $t$, where $t$ is so far only a topological 
time: it faithfully labels instants in a continuous fashion, but 
no implication is made as to its relation to actual clock readings. 
The statement of such relations can eventually only be made on the 
basis of dynamical models for clocks coupled to the gravitational 
field. 

By means of the foliation we now recover a notion of time: we view 
spacetime, $(M,g)$, as the one-parameter family of spaces,
$t\mapsto (\Sigma,h_t)$. Spacetime then becomes nothing but a 
`trajectory of spaces'. In this way we obtain a dynamical system whose 
configuration variable is the Riemannian metric on a 3-manifold 
$\Sigma$. It is to make this point precise that we carefully 
distinguish between the manifold $\Sigma$ and its images 
$\Sigma_t$ in $M$. In the dynamical formulation given now, there 
simply is no spacetime to start with and hence no possibility to 
embed $\Sigma$ into something. Only \emph{after} solving the dynamical 
equations can we construct spacetime and interpret the time 
dependence of the metric of $\Sigma$ as being brought about by 
`wafting' $\Sigma$ through $M$ via a one-parameter family of 
embeddings $\mathcal{E}_t$. But initially there is only a 3-manifold 
$\Sigma$ of some topological type\footnote{It can be shown that 
the Einstein equations do not pose any obstruction to the topology 
of $\Sigma$, that is, solutions exist for \emph{any} topology. 
However, one often imposes additional requirements on the solution.
For example, one may require that there exists a moment of 
time symmetry, which will make the corresponding instant $\Sigma_t$ 
a totally geodesic submanifold of $M$, like e.g. in recollapsing 
cosmological models at the moment of maximal expansion. In this case 
the topology of $\Sigma$ will be severely restricted. In fact, 
most topologies $\Sigma$ will only support geometries that always 
expand or contract somewhere.} and the equations of motion together 
with some suitable initial data. For a fuller discussion we refer 
to the comprehensive work by Isham and 
Kucha\v r~\cite{Isham.Kuchar:1985a,Isham.Kuchar:1985b}. 

\section{The Initial-Value Formulation of GR\index{initial-value formulation of GR}}
Whereas a specified motion of $\Sigma$ through $M$, characterized 
by the family of embeddings~(\ref{eq:SigmaEmbedding}), gives rise 
to a one-parameter family of metrics $h_t$, the converse is not true. 
That is to say, it is not true that \emph{any} one-parameter 
family of metrics $h_t$ of $\Sigma$ can be obtained by finding a 
spacetime $(M,g)$  and a one parameter family of embeddings 
$\mathcal{E}_t$, such that (\ref{eq:SpatialMetric}) holds. 

\begin{figure}[htb]
\centering\epsfig{figure=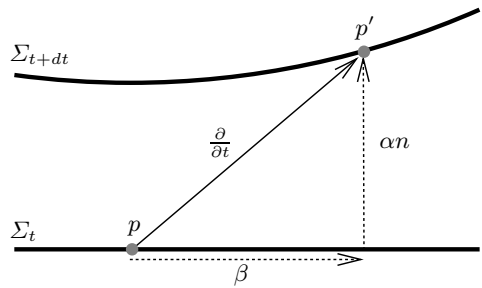,width=0.54\linewidth}
\put(-180,11){\small $\Sigma_{t}$}
\put(-180,78){\small $\Sigma_{t+dt}$}
\put(-135,14){\small $p$}
\put(-49,88){\small $p'$}
\put(-95,-5){\small $\beta$}
\put(-40,45){\small $\alpha n$}
\put(-105,45){\small $\frac{\partial}{\partial t}$}
\caption{\small Infinitesimally nearby leaves $\Sigma_t$ and 
$\Sigma_{t+dt}$. For some point $q\in\Sigma$, the image points 
$p=\mathcal{E}_t(q)$ and $p'=\mathcal{E}_{t+dt}(q)$ are connected 
by the vector $\partial/\partial t\vert_p$, whose components 
tangential and normal to $\Sigma_t$ are $\beta$ 
and $\alpha n$, respectively. $n$ is the normal to $\Sigma_t$
in $M$, $\beta$ is called the `shift vector-field'
and $\alpha$ the `lapse function' on $\Sigma_t$. 
\label{fig:Nico2}}
\end{figure}

Moreover, there is clearly a huge redundancy in creating $(M,g)$ 
from the family $\{(\Sigma,h_t)\mid t\in\mathbb{R}\}$, since 
there are obviously many different motions of $\Sigma$ through 
the same $M$, which give rise to apparently different solution 
curves $h_t$. This redundancy can be locally parameterized by four 
functions, on $\Sigma$: a scalar field $\alpha$ and a vector field 
$\beta$. In the embedding picture, they describe the components 
of the velocity vector field
\begin{equation}
\label{eq:def:d-by-dt}
\frac{\partial}{\partial t}:=\frac{d}{dt}\mathcal{E}_t
\end{equation}
normal and tangential to the leaves $\Sigma_t$ respectively. 
We write 
\begin{equation}
\label{eq:def:LapseShift}
\frac{\partial}{\partial t}=\alpha n+\beta,
\end{equation}
where $n$ is the normal to $\Sigma_t$. The tangential component,
$\beta$, just generates intrinsic diffeomorphisms on each $\Sigma_t$,
whereas the normal component, $\alpha$, really advances one leaf 
$\Sigma_t$ to the next one; see Fig.\,\ref{fig:Nico2}.

For the initial-value problem it is the derivative along the normal 
$n$ of the 3-metric $h$, denoted by $K$, that gives the essential 
information. Hence we write 
\begin{equation}
\label{eq:hDot}
\frac{\partial h_t}{\partial t}=\alpha K_t+L_\beta h_t\,.
\end{equation}
In the embedding picture $K_t$ is the extrinsic curvature of $\Sigma_t$ 
in $M$. 

The first order evolution equations that result from Einstein's field 
equations are then of the general form    
\begin{alignat}{2}
\label{eq:Evolution1}
& \frac{\partial h_t}{\partial t}\,&&=\,F_1(h_t,K_t;\alpha,\beta)\ , \\
\label{eq:Evolution2}
& \frac{\partial K_t}{\partial t}\,&&=\,F_2(h_t,K_t;\alpha,\beta;\text{matter})
\ , 
\end{alignat}
where $F_1$ in (\ref{eq:Evolution1}) is given by the right-hand
side of (\ref{eq:hDot}). $F_2$ is a more complicated 
function whose precise structure need not interest us now and which 
also depends on matter variables; see e.g.~\cite{Giulini:1998}.

\section{Why constraints\index{constraints}}
As we have seen, the initial data for the gravitational variables 
consist of a differentiable 3-manifold $\Sigma$, a Riemannian metric 
$h$ -- the configuration variable, and another symmetric second rank 
tensor field $K$ on $\Sigma$ -- the velocity variable. However, 
the pair $(h,K)$ cannot be chosen arbitrarily. This is because there 
is a large redundancy in describing a fixed spacetime $M$ by a foliation 
(\ref{eq:SigmaEmbedding}). On the infinitesimal level this gauge 
freedom is just the freedom of choosing $\alpha$ and $\beta$.
The gauge transformations generated by $\beta$ are just the spatial 
diffeomorphisms of $\Sigma$. $\beta$ may be an arbitrary function of 
$t$, which corresponds to the fact that we may arbitrarily permute 
the points in each leaf $\Sigma_t$ separately (only restricted by 
some differentiability conditions). The gauge transformations 
generated by $\alpha$ correspond to pointwise changes in the 
velocities with which the leaves $\Sigma_t$ push through $M$. 
These too may vary arbitrarily within the leaves as well as with 
coordinate time $t$.

Whenever there is gauge freedom in a dynamical theory, there 
are so-called \emph{constraints}, that is, conditions which restrict 
the initial data; see e.g.~\cite{Henneaux.Teitelboim:1992}. 
For each gauge freedom parameterized by an arbitrary  function, there 
is one functional combination of the initial data which has to vanish. 
In our case there are four gauge functions, $\alpha$, and the three 
components of $\beta$. Accordingly there are four constraints, which 
group into one \emph{scalar or Hamiltonian constraint},
\index{Hamiltonian constraint}\index{scalar constraint}
 $H[h,K]=0$, and three combined in the 
\emph{vector or diffeomorphism constraint}, $D[h,K]=0$.
\index{diffeomorphism constraint}\index{vector constraint}
Their explicit expressions are:\footnote{Here and below we shall 
write $\sqrt{h}:=\sqrt{\det\{h_{ab}\}}$ and use the abbreviation
$\kappa=8\pi G/c^4$. Hence $\kappa$ has the physical dimension of
$\mathit{s^2\cdot m^{-1}\cdot Kg^{-1}}$. We shall set $c=1$ throughout.} 
\begin{alignat}{2}
\label{eq:HamConstraint1}
& H[h,K] &&\,=\,(2\kappa)^{-1}\,G^{ab\,cd}K_{ab}K_{cd}
-(2\kappa)^{-1}\sqrt{h}\bigl({}^{\scriptscriptstyle (3)}\!R-2\Lambda\bigr)
+\,\sqrt{h}\rho\,,\\
\label{eq:DiffConstraint1}
& D^a[h,K] &&\,=\,-\,\kappa^{-1}\,G^{ab\,cd}\nabla_bK_{cd}+\,\sqrt{h}j^a\, .
\end{alignat}
Here $\rho$ and $j^a$ are the energy- and momentum densities 
of the matter, $\nabla$ and ${}^{\scriptscriptstyle (3)}\!R$ 
are the Levi-Civita connection 
and its associated scalar curvature of $(\Sigma,h)$. 
Finally $G^{ab\,cd}$ is the so called DeWitt metric, \index{DeWitt metric}
which at each point of $\Sigma$ defines an $h$-dependent Lorentzian 
metric on the $1+5$ -- dimensional space of symmetric second-rank 
tensors at that point.\footnote{The Lorentzian signature of the 
DeWitt Metric has nothing to do with the Lorentzian signature 
of the space-time metric: it persists in Euclidean gravity
\index{Euclidean gravity}.} 
Its explicit form is given by 
\begin{equation}
\label{eq:WdWmetric1}
G^{ab\,cd}=
\tfrac{\sqrt{h}}{2}\bigl(h^{ac}h^{bd}+h^{ad}h^{bc}-2h^{ab}h^{cd}\bigr)
\end{equation}
Note that the linear space of symmetric second-rank tensors is 
viewed here as the tangent space (`velocity space') of the space 
$\mathrm{Riem}(\Sigma)$ of Riemannian metrics on $\Sigma$. From 
(\ref{eq:WdWmetric1}) one sees that it is the trace part of the 
`velocities', corresponding to changes of the scale (conformal part) 
of the Riemannian metric, that span the negative-norm velocity 
directions. 

\section{Comparison with conventional form of Einstein's equations}
The presence of constraints and their relation to the evolution 
equations is the key structure in canonical GR. 
It is therefore instructive to point out how this structure arises 
from the conventional, four dimensional form of Einstein's 
equations. Before doing this, it is useful to first remind 
ourselves on the analogous situation in electrodynamics. 

So let us first consider electrodynamics in Minkowski space. 
As usual, we write the field tensor $F$ as exterior differential 
of a vector potential $A$, that is $F=dA$. In components this reads 
$F_{\mu\nu}=\partial_\mu A_\nu-\partial_\nu A_\mu$.
Here $E_i=F_{0i}$ are the components of the electric, 
$B_i=-F_{jk}$ of the magnetic field, where $ijk$ is a cyclic 
permutation of 123. The homogeneous Maxwell equations now simply 
read $dF=0$, whereas the inhomogeneous Maxwell equations 
\index{Maxwell equations}are given by (in components): 
\begin{equation}
\label{eq:InhMaxwellEq}
M^\mu:=\partial_\nu F^{\mu\nu}+\tfrac{4\pi}{c}j^\mu=0\,,
\end{equation}
where here $j^\mu$ is the electric four-current. Due to its 
antisymmetry, the field tensor obeys the identity 
\begin{equation}
\label{eq:MaxwellId1}
\partial_\mu\partial_\nu F^{\mu\nu}\equiv 0\,.
\end{equation}
Taking the divergence of (\ref{eq:InhMaxwellEq}) and using 
(\ref{eq:MaxwellId1}) leads to 
\begin{equation}
\label{eq:MaxwellId2}
\partial_\mu M^\mu\equiv\tfrac{4\pi}{c}\partial_\mu j^\mu=0\,,
\end{equation}
showing the well known fact that Maxwell's equations imply 
charge conservation as integrability condition. 

Let us now interpret the role of charge conservation in the 
initial-value problem. Decomposing (\ref{eq:MaxwellId1}) into 
space and time derivatives yields
\begin{equation}
\label{eq:MaxwellId3}
\partial_0\partial_\nu F^{0\nu}\equiv 
-\partial_a\partial_\nu F^{a\nu}\,.
\end{equation}
Even though the right-hand side contains third derivatives in the 
field $A_\mu$, time derivatives appear at most in second order
(since $\partial_a$ is spatial). Hence, since it is an identity, 
$\partial_\nu F^{0\nu}$ contains time derivatives only up to first 
order. But the initial data for the second order equation 
(\ref{eq:InhMaxwellEq}) consist of the field $A_\mu$ and its 
first time derivative. Hence the time component $M^0$ of Maxwell's 
equations gives a relation amongst initial data, in other words, 
it is a \emph{constraint}. Clearly this is just the Gau{\ss} constraint
\index{Gau{\ss} constraint} 
$\vec\nabla\cdot\vec E-4\pi\rho=0$ (here $\rho$ is the electric 
charge density). Only the three spatial components of 
(\ref{eq:InhMaxwellEq}) contain second time derivatives and hence 
propagate the fields. They provide the evolutionary part of
Maxwell's equations. 

Now, assume we are given initial data satisfying the constraint 
$M^0=0$, which we evolve according to $M^a=0$. How can we be sure 
that the evolved data again satisfy the constraint? To see when 
this is the case, we use the identity (\ref{eq:MaxwellId2}) and 
solve it for the time derivative of $M^0$: 
\begin{equation}
\label{eq:MaxwellId4}
\partial_0 M^0\equiv 
-\partial_a M^a+\tfrac{4\pi}{c}\partial_\mu j^\mu\,.
\end{equation}
This shows: if initially $M^a=0$ (and hence $\partial_a M^a=0$),
then the constraint $M^0=0$ is preserved in time if and only if 
$\partial_\mu j^\mu=0$. Charge conservation is thus recognized as 
the necessary and sufficient condition for the compatibility 
between the constraint part and the evolutionary part of Maxwell's 
equations.    

Finally we wish to make another remark concerning the interplay 
between constraints and evolution equations. It is clear that 
a solution $F^{\mu\nu}$ to (\ref{eq:InhMaxwellEq}) satisfies the constraint 
on \emph{any} simultaneity hypersurface of an inertial observer
(i.e. spacelike plane). If the normal to the hypersurface is 
$n_\mu$, this just states that $M^\mu=0$ implies $M^\mu n_\mu=0$. 
But the converse is obviously also true: if  $M^\mu n_\mu=0$ for 
all timelike $n_\mu$, then $M^\mu=0$. In words: given an 
electromagnetic field that satisfies the constraint (for given 
external current $j^\mu$) on \emph{any} spacelike plane in 
Minkowski space, then this field must necessarily satisfy 
Maxwell's equations. In this sense, Maxwell's equations are 
the \emph{unique} propagation law that is compatible with 
Gau{\ss}' constraint. \index{Gau{\ss} constraint}

After this digression we return to GR, where we 
can perform an entirely analogous reasoning. We start with Einstein's
equations, in which the spacetime metric $g_{\mu\nu}$ is the 
analog of $A_\mu$ and the Einstein tensor 
$G^{\mu\nu}:=R^{\mu\nu}-\tfrac{1}{2}g^{\mu\nu}R$ is the 
analog of $\partial_\nu F^{\mu\nu}$. They read  
\begin{equation}
\label{eq:EinsteinEquation}
E^{\mu\nu}:=G^{\mu\nu}-\Lambda-\kappa T^{\mu\nu}=0\,.
\end{equation}
Due to four dimensional diffeomorphism invariance, we have the 
identity (twice contracted second Bianchi-Identity): 
\begin{equation}
\label{eq:BianchiId}
\nabla_\mu G^{\mu\nu}\equiv 0\,,
\end{equation}
which is the analog of (\ref{eq:MaxwellId1}). Taking the covariant 
divergence of (\ref{eq:EinsteinEquation}) and using 
(\ref{eq:BianchiId}) yields 
\begin{equation}
\label{eq:EinsteinId}
\nabla_\mu E^{\mu\nu}=-\kappa\nabla_\mu T^{\mu\nu}=0\,,
\end{equation}
which is the analog of  (\ref{eq:MaxwellId2}). Hence the vanishing 
covariant divergence of $T^{\mu\nu}$ is an integrability condition 
of Einstein's equations, just as the divergencelessness of the 
electric four-current was an integrability condition of Maxwell's 
equations.\footnote{There is, however, a notable difference in 
the physical interpretation of divergencelessness of a tensor 
field on one hand, and a vector field on the other:
$\nabla_\mu T^{\mu\nu}=0$ does not as such imply a conservation 
law. Only in presence of a spacetime symmetry, i.e. a Killing 
vector field $K_\nu$, the current $J^\mu=T^{\mu\nu}K_\nu$ is 
conserved, $\nabla_\mu J^\mu=0$, and hence gives rise to a conserved 
quantity.} 

In order to talk about `evolution', we consider the foliation 
(\ref{eq:SigmaEmbedding}) of $M$ and locally use coordinates 
$\{x^0,x^a\}$ such that $\partial/\partial x^0$ is the normal 
$n$ to the leaves and all $\partial/\partial x^a$ are tangential. 
Expanding (\ref{eq:BianchiId}) in terms of partial derivatives 
gives:
\begin{equation}
\label{eq:BianchiIdExp}
\partial_0G^{0\nu}=-\partial_aG^{a\nu}
-\Gamma^\mu_{\mu\lambda}G^{\lambda\nu}
-\Gamma^\nu_{\mu\lambda}G^{\mu\lambda}\,,
\end{equation}
which is the analog of (\ref{eq:MaxwellId3}).
Now, since the $G^{\mu\nu}$ contain at most second and the 
$\Gamma_{\mu\nu}^\lambda$ at most first derivatives of the 
metric $g_{\mu\nu}$, this identity immediately shows that the 
four components $G^{0\nu}$ ($\nu=0,1,2,3$) contain at most first 
time derivatives $\partial/\partial x^0$. But Einstein's 
equations are of second order, hence the four equations 
$E^{0\nu}=0$ are relations amongst the initial data, rather 
than being evolution equations. In fact, up to a factor of -2 
they are just the 
constraints~(\ref{eq:HamConstraint1}--\ref{eq:DiffConstraint1}):
\begin{alignat}{3}
\label{eq:HamConstraint2}
& H   &&\,=\,-2E^{00} &&\,=\,-2( G^{00}-\Lambda-\kappa T^{00})\,,\\
\label{eq:DiffConstraint2}
& D^a &&\,=\,-2E^{0a} &&\,=\,-2(G^{0a}-\Lambda-\kappa T^{0a})\,.
\end{alignat}
Moreover, the remaining purely spatial components of Einstein's 
equations are equivalent to the twelve first-order evolution 
equations (\ref{eq:Evolution1}-\ref{eq:Evolution2}).     

The interplay between constraints and evolution equations can 
now be followed along the very same lines as for the electrodynamic 
analogy. Expanding the left equality of (\ref{eq:EinsteinId}) in 
terms of partial derivatives gives
\begin{equation}
\label{eq:EinsteinIdExp}
\partial_0E^{0\nu}=-\partial_aE^{a\nu}
-\Gamma^\mu_{\mu\lambda}E^{\lambda\nu}
-\Gamma^\nu_{\mu\lambda}E^{\mu\lambda}
-\kappa\nabla_\mu T^{\mu\nu}\,,
\end{equation}
which is the analog of (\ref{eq:MaxwellId4}). It shows that the 
constraints are preserved by the evolution if and only if the 
energy-momentum tensor of the matter has vanishing covariant 
divergence. 

Let us now turn to the last analogy: the uniqueness of the 
evolution preserving constraints. Clearly Einstein's equations 
$E^{\mu\nu}$ imply $E^{\mu\nu}n_\mu=0$ for any timelike vector 
field $n_\mu$. Hence the constraints are satisfied on any spacelike 
slice through spacetime. Again the converse is also true: given
a gravitational field such that $E^{\mu\nu}n_\mu=0$ for any 
timelike $n_\mu$ (and given external $T^{\mu\nu}$), then this 
field must necessarily satisfy Einstein's equations. In this 
sense Einstein's equations follow uniquely from the condition of 
constraint preservation. 

This property will be crucial for the interpretation of the
quantum theory discussed below. We know from quantum mechanics that the
classical trajectories have completely disappeared at the fundamental
level. As we have discussed above, the analogue to a trajectory
is in GR provided by a spacetime 
given as a set of three dimensional geometries.
In quantum gravity, the spacetime will therefore disappear like the
classical trajectory in quantum mechanics. It is therefore not 
surprising that the evolution equations \eqref{eq:Evolution1} and 
\eqref{eq:Evolution2} will be absent in quantum gravity. All the 
information will be contained in the quantized form of the constraints 
\eqref{eq:HamConstraint1} and \eqref{eq:DiffConstraint1}.

\section{Canonical gravity\index{Canonical gravity}}
We have seen above that Einstein's equations can be written as a 
dynamical system (\ref{eq:Evolution1}--\ref{eq:Evolution2})
with constraints (\ref{eq:HamConstraint1}--\ref{eq:DiffConstraint1}).
Here we wish to give its canonical formulation. Basically this 
means to introduce momenta for the velocities and write the 
first-order equations of motions as Hamilton equations. This means to 
identify the Poisson structure and the Hamiltonian. The result is 
this: As before, the configuration variable is the Riemannian 
metric $h_{ab}$ on $\Sigma$. Its canonical momentum is now given by
\begin{equation}
\label{eq:CanMomentum}
\pi^{ab}=(2\kappa)^{-1}\,G^{ab\,cd}K_{cd}
        =(2\kappa)^{-1}\,\sqrt{h}(K^{ab}-h^{ab}K_c^c)\,,
\end{equation}
%

so that the Poisson brackets are
\begin{equation}
\label{eq:PoissonBrackets}
\{h_{ab}(x),\pi^{cd}(y)\}
=\tfrac{1}{2}(\delta^c_a\delta^d_b+\delta^d_a\delta^c_b)\delta^{(3)}(x,y)\,, 
\end{equation}
where $\delta^{(3)}(x,y)$ is the Dirac distribution on $\Sigma$. 

Elimination of $K_{ab}$ in favour of $\pi^{ab}$ in the constraints 
leads to their canonical form: 
\begin{alignat}{2}
\label{eq:HamConstraint3}
& H[h,\pi] &&\,=\,2\kappa\,G_{ab\,cd}\pi^{ab}\pi^{cd}
\,-\,(2\kappa)^{-1}\sqrt{h}({}^{\scriptscriptstyle (3)}\!R-2\Lambda)\,
+\,\sqrt{h}\rho\,,\\
\label{eq:DiffConstraint3}
& D^a[h,\pi] &&\,=-2\nabla_b\pi^{ab}\,+\,\sqrt{h}j^a\,,
\end{alignat}
where now\footnote{Note the difference in the factor of 
two in the last term, as compared to (\ref{eq:WdWmetric1}). 
$G_{ab\,cd}$ is the inverse to $G^{ab\,cd}$, i.e. 
$G^{ab\,nm}G_{nm\,cd}=\tfrac{1}{2}(\delta^a_c\delta^b_d+
\delta^a_d\delta^b_c)$, and \emph{not} obtained by simply 
lowering the indices using $h_{ab}$.}
\begin{equation}
\label{eq:WdWMetric2}
G_{ab\,cd}=\tfrac{1}{2\sqrt{h}}(h_{ac}h_{bd}+h_{ad}h_{bc}-h_{ab}h_{cd})\,.
\end{equation}
Likewise, rewriting (\ref{eq:Evolution1}--\ref{eq:Evolution2})
in terms of the canonical variables shows that they are just 
the flow equations for the following Hamiltonian: 
\begin{equation}
\label{eq:Superhamiltonian}
\mathcal{H}[h,\pi]=\int_\Sigma d^3x
\bigl\{\alpha(x) H[h,\pi_{ab}](x)+\beta^a(x)D_a[h,\pi](x)\bigr\}+\text{boundary terms}\,.
\end{equation}
The crucial observation to be made here is, that, up to boundary terms, 
the total Hamiltonian is a combination of pure constraints. The boundary 
terms generally appear if $\Sigma$ is non-compact, as it will be the 
case for the description of isolated systems, like stars or black holes. 
In this case the boundary terms are taken over closed surfaces at 
spatial infinity and represent conserved Poincar\'e charges, like energy, 
linear- and angular momentum, and the quantity associated with asymptotic 
boost transformations. If, however, $\Sigma$ is closed (i.e. compact 
without boundary) all of the evolution will be generated by constraints, 
that is, pure gauge transformations! In that case, evolution, as 
described here, is not an observable change. For that to be the case 
we would need an extrinsic clock, with respect to which `change' can 
be defined. But a closed universe already contains -- by definition -- 
everything physical, so that no external clock exists. Accordingly, 
there is no external time parameter. Rather, all physical time 
parameters are to be constructed from within our system, that is, as 
functional of the canonical variables. A priori there is no preferred 
choice of such an intrinsic time parameter. The absence of an 
extrinsic time and the non-preference of an intrinsic one is commonly 
known as the \emph{problem of time}\index{problem of time} in Hamiltonian
(quantum-) cosmology.  

Finally we turn to the commutation relation between the various 
constraints. For this it is convenient to integrate the local 
constraints (\ref{eq:HamConstraint3}--\ref{eq:DiffConstraint3})
over lapse and shift functions. Hence we set (suppressing the 
phase-space argument $[h,\pi]$) 
\begin{alignat}{2}
\label{eq:SmearedHConstraint}
& \mathcal{H}(\alpha)&&\,=\,\int_\Sigma d^3x\,H(x)\,\alpha(x)\,,\\
\label{eq:SmearedDConstraint}
& \mathcal{D}(\beta) &&\,=\,\int_\Sigma d^3x\,D^a(x)\,\beta_a(x)\,.
\end{alignat}
A straightforward but slightly tedious computation gives
\begin{alignat}{2}
\label{eq:CAlgebra1}
& \{\mathcal{D}(\beta),\mathcal{D}(\beta')\}
&&\,=\,\mathcal{D}([\beta,\beta'])\,,\\
\label{eq:CAlgebra2}
& \{\mathcal{D}(\beta),\mathcal{H}(\alpha)\}
&&\,=\,\mathcal{H}(\beta(\alpha))\,,\\
\label{eq:CAlgebra3}
& \{\mathcal{H}(\alpha),\mathcal{H}(\alpha')\}
&&\,=\,\mathcal{D}(\alpha\nabla\alpha'-\alpha'\nabla\alpha)\,.
\end{alignat}
There are three remarks we wish to make concerning these relations. 
First,~(\ref{eq:CAlgebra1}) shows that the diffeomorphism generators
form a Lie subalgebra. Second,~(\ref{eq:CAlgebra2}) shows that this 
Lie subalgebra is not a Lie ideal. This means that the flow of the 
Hamiltonian constraint does not leave invariant the constraint 
hypersurface of the diffeomorphism constraint. 
Finally, the term $\alpha\nabla\alpha'-\alpha'\nabla\alpha$ in 
(\ref{eq:CAlgebra3}) contains the canonical variable $h$, which is 
used implicitly to raise the index in the differential in order to 
get the gradient $\nabla$. This means that the relations above 
do not make the set of all $\mathcal{H}(\alpha)$ and all 
$\mathcal{D}(\beta)$ into a Lie algebra.\footnote{Sometimes this 
is expressed by saying that this is an `algebra with structure 
functions'.} 

\section{The general kinematics of hypersurface deformations}
In this section we wish to point out that the relations 
(\ref{eq:CAlgebra1}--\ref{eq:CAlgebra3}) follow a general pattern,
namely to represent the `algebra' of hypersurface deformations, 
or in other words, infinitesimal changes of embeddings 
$\mathcal{E}:\Sigma\rightarrow M$. To make this explicit, 
we introduce local coordinates $x^a$ on $\Sigma$ and $y^\mu$
on $M$. An embedding is then locally given by four functions 
$y^\mu(x)$, such that the $3\times 4$ matrix $y^\mu_{,a}$ has
its maximum rank 3 (we write $y^\mu_{,a}:=\partial_ay^\mu$). 
The components of the normal to the image 
$\mathcal{E}(\Sigma)\subset M$ are denoted by $n^\mu$, which 
should be considered as functional of $y^\mu(x)$. The generators 
of normal and tangential deformations of $\mathcal{E}$ with 
respect to the lapse function $\alpha$ and shift vector field 
$\beta$ are then given by 
\begin{alignat}{2}
\label{eq:NormalDef}
& N_\alpha &&\,=\,
\int_\Sigma d^3x\ \alpha(x)\,n^\mu[y(x)]\,\frac{\delta}{\delta y^\mu(x)}\,,\\ 
\label{eq:TangentDef}
& T_\beta &&\,=\,
\int_\Sigma d^3x\ \beta^a(x)\,y^\mu_{,a}(x)\, 
\frac{\delta}{\delta y^\mu(x)}\,,
\end{alignat}
which may be understood as tangent vectors to the space of 
embeddings of $\Sigma$ into $M$. 
A calculation\footnote{Equation (\ref{eq:AlgHypDef1}) is 
immediate. To verify (\ref{eq:AlgHypDef2}--\ref{eq:AlgHypDef3}) 
one needs to compute $\delta n^{\mu}[y(x)]/\delta y^{\nu}(x')$.
This can be done in a straightforward way by varying 
\[
g(y(x))_{\mu\nu}n^\mu[y(x)] n^\nu[y(x)]=-1
\quad\text{and}\quad
g_{\mu\nu}(y(x))\,y^\mu_{,a}(x)\,n^\nu[y(x)]=0
\]
with respect to $y(x)$.} then leads to the following 
commutation relations 
\begin{alignat}{2}
\label{eq:AlgHypDef1}
& [T_\beta\,,\,T_{\beta'}]   &&\,=\,-\, T_{[\beta,\beta']}\,,\\
\label{eq:AlgHypDef2}
& [T_\beta\,,\,N_\alpha]     &&\,=\,-\, N_{\beta(\alpha)} \,,\\ 
\label{eq:AlgHypDef3}
& [N_\alpha\,,\,N_{\alpha'}] &&\,=\,-\, 
T_{\alpha\nabla\alpha'-\alpha'\nabla\alpha}\,.
\end{alignat}
Up to the minus signs this is just 
(\ref{eq:CAlgebra1}-\ref{eq:CAlgebra3}). The minus signs are just 
the usual ones that one always picks up in going from the action of 
vector fields to the Poisson action of the corresponding phase-space 
functions. (In technical terms, the mapping from vector fields to 
phase-space functions is a Lie-\emph{anti}-homomorphism.)

This shows that (\ref{eq:CAlgebra1}--\ref{eq:CAlgebra3}) just mean
that we have a Hamiltonian realization of hypersurface deformations. 
In particular, (\ref{eq:CAlgebra1}--\ref{eq:CAlgebra3}) is neither
characteristic of the action nor the field content: \emph{Any} 
four dimensional diffeomorphism invariant theory will gives rise 
to this very same `algebra'. It can be shown that under certain 
general locality assumptions the expressions (\ref{eq:HamConstraint3}) 
and (\ref{eq:DiffConstraint3}) give the unique 2-parameter (here 
$\kappa$ and $\Lambda$) family of realizations for $N$ and $T$
satisfying (\ref{eq:AlgHypDef1}--\ref{eq:AlgHypDef3}) on the phase 
space parameterized by $(h_{ab}\,,\,\pi^{ab})$; 
see~\cite{Hojman.etal:1976} and also~\cite{Kuchar:1974}.

\section{Topological issues}
As we have just discussed, Einstein's equations take the form of a 
constrained Hamiltonian system if put into canonical form.
 The unconstrained configuration space 
is the space of all Riemannian metrics on some chosen 3-manifold 
$\Sigma$. This space is denoted by $\text{Riem}(\Sigma)$. Any two 
Riemannian metrics that differ by an action of the diffeomorphism
constraint are gauge equivalent and hence to be considered as 
physically indistinguishable. Let us briefly mention that the 
question of whether and when the diffeomorphism constraint actually 
generates all diffeomorphisms of $\Sigma$ is rather subtle. Certainly, 
what is generated lies only in the identity component of the latter, 
but even on that it may not be onto. This occurs, for example, in 
the case where $\Sigma$ contains asymptotically flat ends with 
non-vanishing Poincar\'e charges associated. Asymptotic Poincar\'e
transformations are then not interpreted as gauge transformations
(otherwise the Poincar\'e charges were necessarily zero), but as 
proper physical symmetries (i.e. changes of state that are observable
in principle).

\begin{figure}[htb]
\centering\epsfig{figure=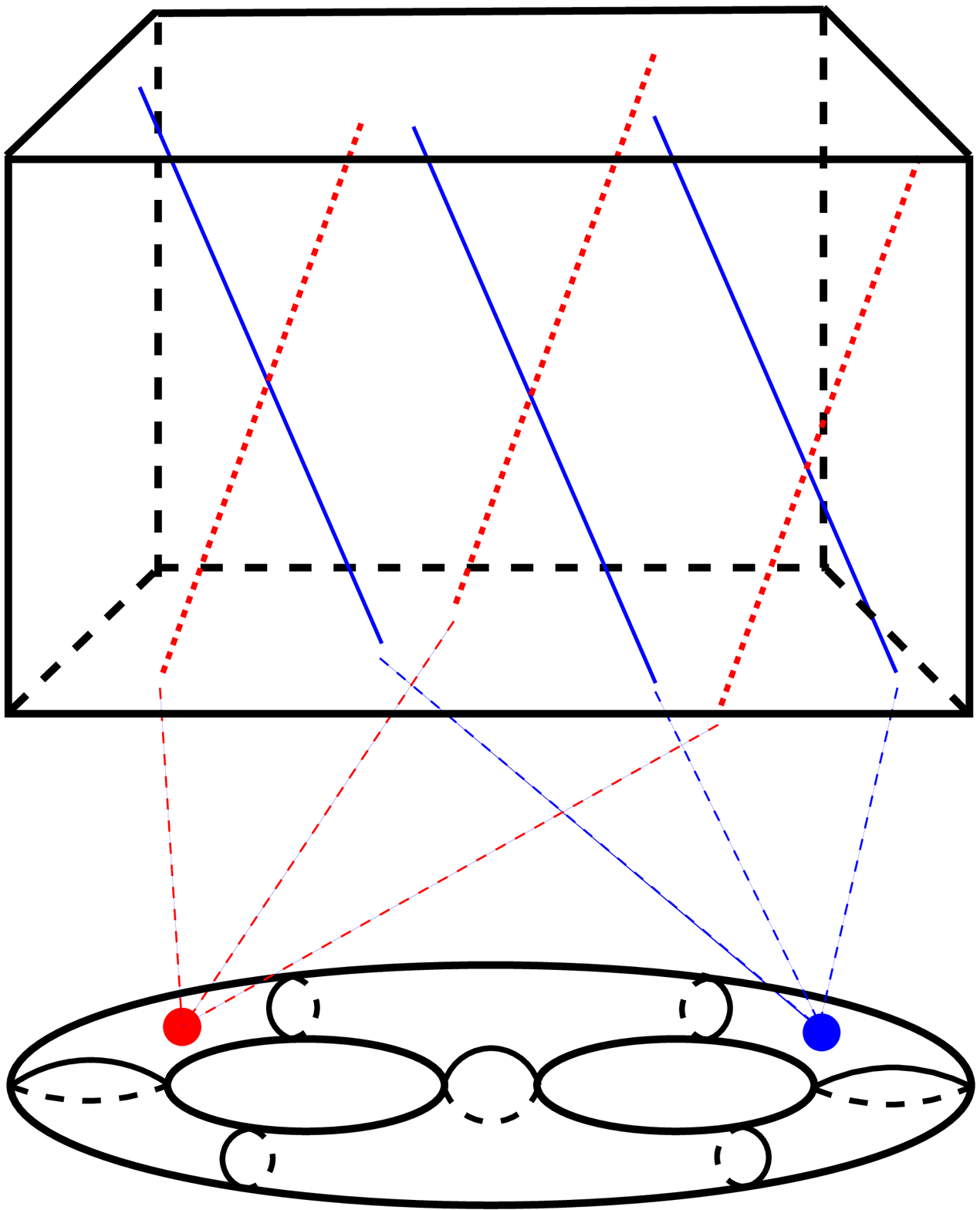,width=0.45\linewidth}
\put(5,130){$\text{Riem}(\Sigma)$}
\put(5,18){$\mathcal{S}(\Sigma)$}
\caption{\small The topologically trivial space $\text{Riem}(\Sigma)$,
here drawn as the box above, is fibered by the action of the 
diffeomorphism group. The fibers are the straight lines in the 
box, where the sets consisting of three dashed and three solid 
lines, respectively, form one fiber each. In the quotient space 
$\mathcal{S}(\Sigma)$ each fiber is represented by one point only. 
By taking the quotient, $\mathcal{S}(\Sigma)$ receives the 
non-trivial topology from $\text{Diff}(\Sigma)$. To indicate this, 
$\mathcal{S}(\Sigma)$ is represented as a double torus.  
\label{fig:Nico3}}
\end{figure}

Leaving aside the possible difference between what is generated by 
the constraints and the full group $\text{Diff}(\Sigma)$ of 
diffeomorphisms of $\Sigma$, we may consider the quotient space 
$\text{Riem}(\Sigma)/\text{Diff}(\Sigma)$ of Riemannian 
\emph{geometries}. This space is called \emph{superspace} in the 
relativity community (this has nothing to do with supersymmetry), 
which we denote by $\mathcal{S}(\Sigma)$. 
Now from a topological viewpoint $\text{Riem}(\Sigma)$ is rather 
trivial. It is a cone\footnote{Any real positive multiple $\lambda h$ of 
$h\in\text{Riem}(\Sigma)$ is again an element of $\text{Riem}(\Sigma)$.}
in the (infinite dimensional) vector space of all symmetric second-rank 
tensor fields. But upon factoring out $\text{Diff}(\Sigma)$ the quotient 
space $\mathcal{S}(\Sigma)$ inherits some of the topological information 
concerning $\Sigma$, basically because $\text{Diff}(\Sigma)$ contains 
that information~\cite{Giulini:1995a}. This is schematically drawn in 
Fig.\,\ref{fig:Nico3}.

In a certain generalized sense, GR is a dynamical 
system on the phase space (i.e. cotangent bundle) built over 
superspace. The topology of superspace is characteristic for the 
topology of $\Sigma$, though in a rather involved way. 
Note that, by construction, the Hamiltonian evolution is that of 
a varying embedding of $\Sigma$ into spacetime. Hence the images 
$\Sigma_t$ are all of the same topological type. This is why 
canonical gravity in the formulation given here cannot describe 
transitions of topology. 

\begin{figure}[htb]
\begin{minipage}[b]{0.32\linewidth}
\centering\epsfig{figure=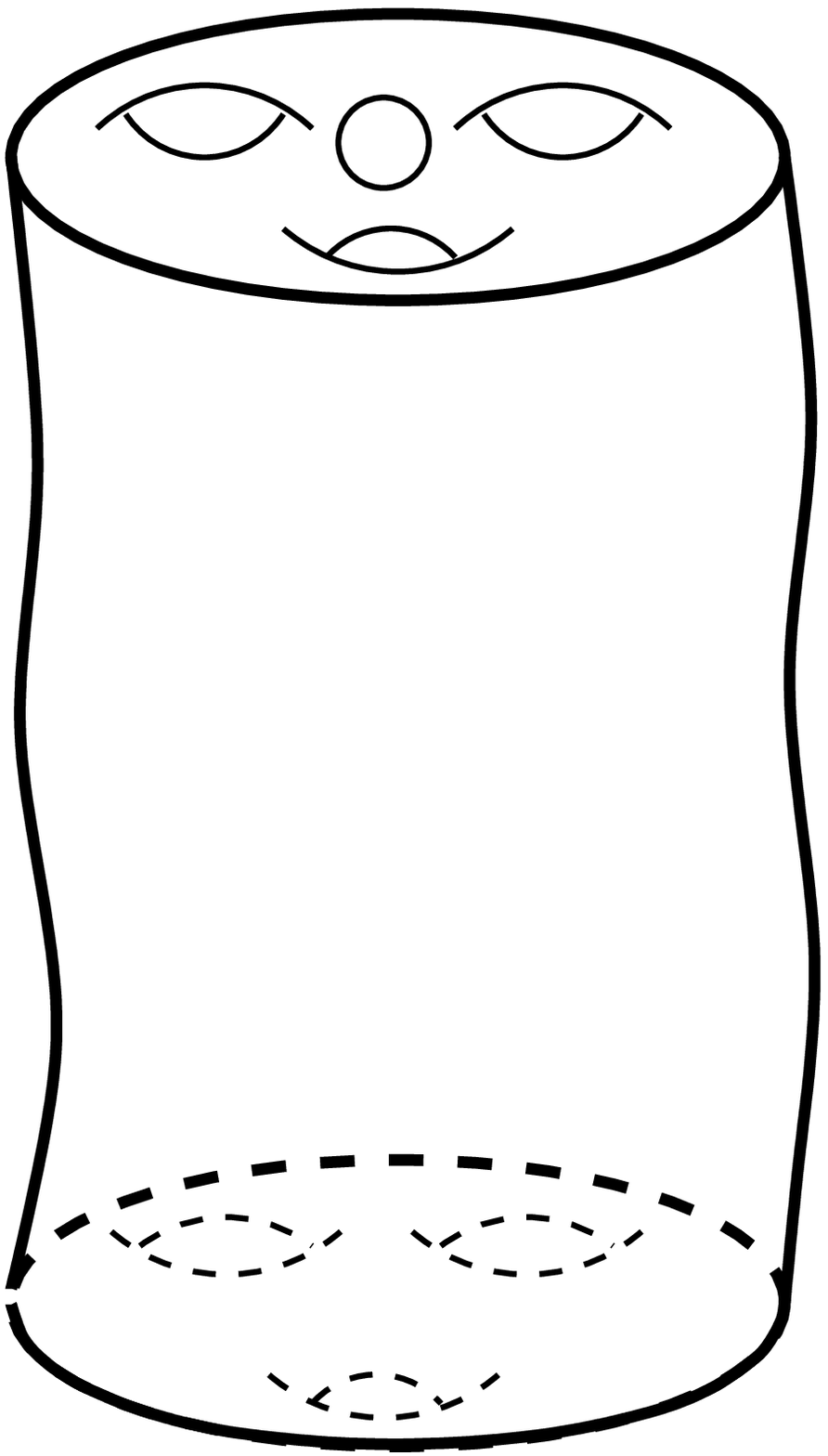,width=1.0\linewidth}
\end{minipage}
\hfill
\begin{minipage}[b]{0.44\linewidth}
\centering\epsfig{figure=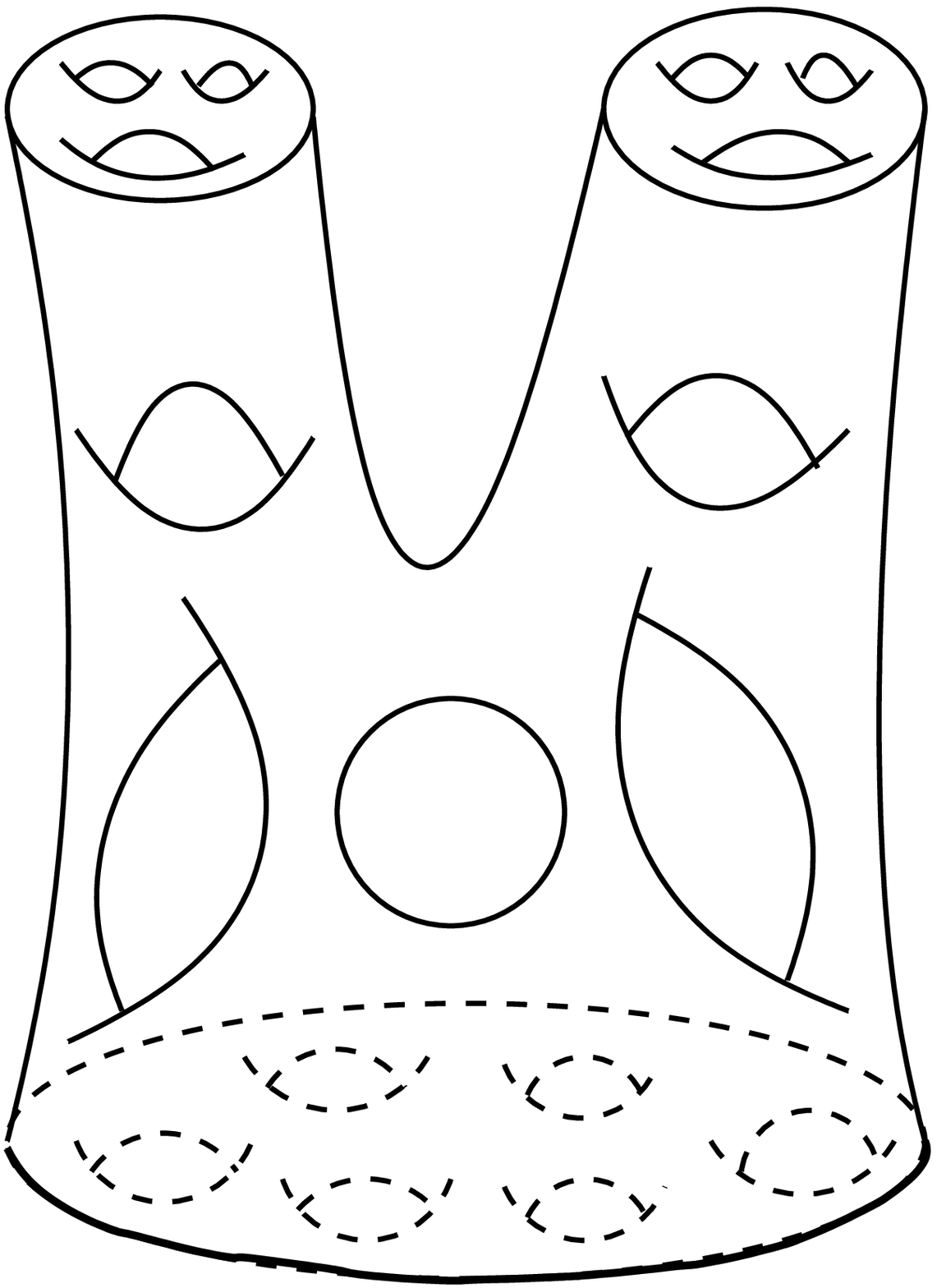,width=1.0\linewidth}
\end{minipage}
\put(-262,14){$\Sigma_{\rm i}$}
\put(-260,160){$\Sigma_{\rm f}$}
\put(-290,85){$M$}
\put(-47,20){$\Sigma_{\rm i}$}
\put(-64,190){\small $\Sigma_{\rm f}^1$}
\put(-95,190){\small $\Sigma_{\rm f}^2$}
\put(-82,100){$M$}
\caption{\small Spacetimes in which spatial sections change topology.
In the left picture the initial universe $\Sigma_{\rm i}$ has three, 
the final $\Sigma_{\rm f}$ four topological features (`holes') -- 
it `grows a nose' while staying connected. In the right 
picture the initial universe $\Sigma_i$ splits into two 
copies $\Sigma_{\rm f}^{1{,}2}$, so that 
$\Sigma_f=\Sigma_{\rm f}^1\cup\Sigma_{\rm f}^2$. 
In both cases, the interpolating spacetime $M$ can be chosen to 
carry a Lorentzian metric with respect to which initial and final 
hypersurfaces are spacelike, possibly at the price of making 
$M$ topologically complicated, like indicated in the right 
picture. 
\label{fig:Nico4}}
\end{figure}

Note, however, that this is not at all an implication by Einstein's
equations. Rather, it is a consequence of our restriction to 
spacetimes that admit a global spacelike foliation. There are many 
solutions to Einstein's equations that do not admit such foliations
globally. This means, that these spacetimes cannot be constructed 
by integrating the equations of motions 
(\ref{eq:Evolution1}--\ref{eq:Evolution2}) successively from some 
initial data. Should we rule out all other solutions? The general 
feeling seems to be, that at least in quantum gravity, topology 
changing classical solutions should not be ruled out as possible 
contributors in the sum over histories (path integral). 
Fig.\,\ref{fig:Nico4} shows two such histories. Whereas in the left 
picture the universe simply `grows a nose', it bifurcates in the 
right example to become disconnected. 

One may ask whether there are topological restrictions to 
such transitions. First of all, it is true (though not at 
all obvious) that for any given two 3-manifolds 
$\Sigma_{\rm i}$, $\Sigma_{\rm f}$  (neither needs to be 
connected) there is a 4-manifold $M$ whose boundary is just 
$\Sigma_{\rm i}\cup\Sigma_{\rm f}$. In fact, there are 
infinitely many such $M$. Amongst them, one can always find 
some which can be endowed with a globally regular Lorentz metric 
$g$, such that $\Sigma_{\rm i}$ and $\Sigma_{\rm f}$ are 
spacelike. However, if topology changes, $(M,g)$ necessarily contains 
closed timelike curves~\cite{Geroch:1967}. This fact has sometimes been 
taken as rationale for ruling out topology change in (classical) 
GR. But it should be stressed that closed 
timelike curves do not necessarily ruin conventional concepts 
of predictability. In any case, let us accept this slight 
pathology and ask what other structures we wish to define on $M$. 
For example, in order to define fermionic matter fields on $M$ 
we certainly wish to endow $M$ with a $SL(2,\mathbb{C})$ spin 
structure. This is where now the first real obstructions for 
topological transitions 
appear~\cite{Gibbons.Hawking:1992}.\footnote{Their result is the 
following: Let $\Sigma=\Sigma_{\rm i}\cup\Sigma_{\rm f}$ be the 
spacelike boundary of the Lorentz manifold $M$, then 
$\dim\bigl(H^0(\Sigma,\mathbf{Z}_2)\bigr)
+\dim\bigl(H^1(\Sigma,\mathbf{Z}_2)\bigr)$ has to be even for 
$M$ to admit an $SL(2,\mathbb{C})$ spin  structure.} 
It is then possible to translate them into selection rules 
for transitions between all known 3-manifolds~\cite{Giulini:1992}. 

So far the considerations were purely kinematical. What additional 
obstructions arise if the spacetime $(M,g)$ is required to satisfy 
the field equations? Here the situation becomes worse. It is, for example,
known that any topology-changing spacetime that satisfies Einstein's 
equations with matter that satisfies the weak-energy condition 
$T_{\mu\nu}l^\mu l^\nu\geq 0$ for all lightlike $l^\mu$ must 
necessarily be singular.\footnote{In fact, this result can be 
considerably strengthened: Instead of invoking Einstein's equations 
we only need to require $R_{\mu\nu}l^\mu l^\nu\geq 0$ for all 
lightlike $l^\mu$.} Hence it seems that we need to consider 
degenerate metrics already on the classical level if topology 
change is to occur. Can we relax the notion of 
`solution to Einstein's equations' so as to contain these degenerate 
cases as well? The answer is `yes' if instead of taking the metric
as basic variable we rewrite the equations in terms of vierbeine
and connections (first oder formalism). It turns out that the kind 
of singularities one has to cope with are very mild indeed:
the vierbeine  become degenerate on sets of measure zero but, 
somewhat surprisingly, the curvature stays bounded everywhere.
In fact, there is a very general method to generate an abundance 
of such solutions~\cite{Horowitz:1991}. 

It is a much debated question whether topology changing amplitudes 
are suppressed or, to the contrary, needed in quantum gravity. 
On one hand, it has been shown in the context of specific lower 
dimensional models that matter fields on topology-changing 
backgrounds may give rise to singularities corresponding to infinite 
densities of particle production~\cite{Anderson.DeWitt:1986}. 
On the other hand, leaving out topology changing amplitudes in the 
sum-over-histories approach is heuristically argued to be in 
conflict with expected properties of localized 
pseudo-particle-like excitations in gravity (so called geons), 
like, for example, the usual spin-statistic relation~\cite{Sorkin:1997}. 
Here there still seems to be much room for speculations.

\section{Geometric issues}
Just in the same way as any Lagrangian theory endows the configuration 
space with the kinetic-energy metric, $\text{Riem}(\Sigma)$ inherits a 
metric structure from the `kinetic-energy' part of 
(\ref{eq:HamConstraint1}). Tangent vectors at $h\in\text{Riem}(\Sigma)$
are symmetric second-rank tensor fields on $\Sigma$ and their inner 
product is given by the so-called \emph{Wheeler--DeWitt metric}:
\index{Wheeler--DeWitt metric}
\begin{equation}
\label{eq:DeWittMetric}
\mathcal{G}_h(V,V')=\int_\Sigma d^3x\,G^{ab\,cd}V_{ab}V'_{cd}\,.
\end{equation}
Due to the pointwise Lorentzian signature (1+5) of $G^{ab\,cd}$ it
is of a hyper-Lorentzian structure with infinitely many negative, null, 
and positive directions each. However, not all directions in the 
tangent space $T_h(\text{Riem}(\Sigma))$ correspond to physical 
changes. Those generated by diffeomorphism, which are of the 
form $V_{ab}=\nabla_a \beta_b+\nabla_b \beta_a$ for some vector 
field $\beta$ on $\Sigma$ are pure gauge. We call them \emph{vertical}.
\begin{figure}[htb]
\centering\epsfig{figure=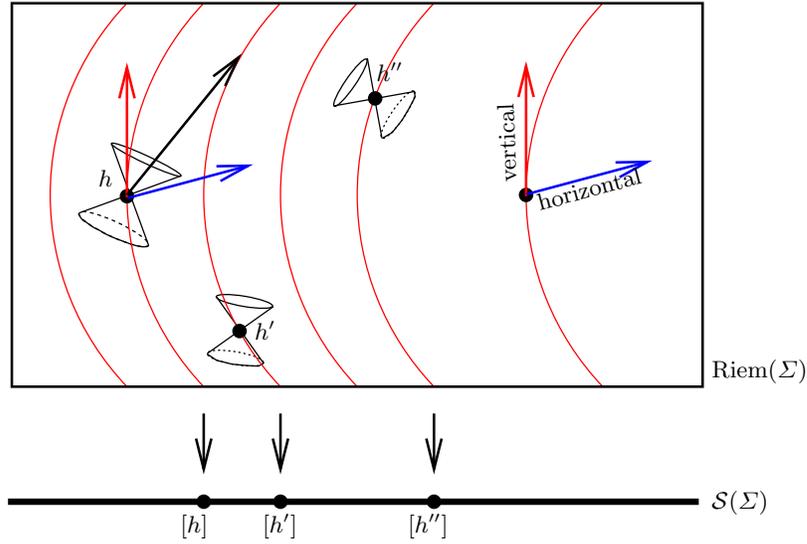,width=0.8\linewidth}
\put(2,50){$\text{Riem}(\Sigma)$}
\put(2,0){$\mathcal{S}(\Sigma)$}
\put(-72,124){\footnotesize\begin{rotate}{92}vertical\end{rotate}}
\put(-63,113){\footnotesize\begin{rotate}{15}horizontal\end{rotate}}
\put(-199,-9){\footnotesize $[h]$}
\put(-168,-9){\footnotesize $[h']$}
\put(-113,-9){\footnotesize $[h'']$}
\put(-230,122){\footnotesize $h$}
\put(-171,64){\footnotesize $h'$}
\put(-125,162){\footnotesize $h''$}
\caption{\small The space $\text{Riem}(\Sigma)$, fibered by the orbits 
of $\text{Diff}(\Sigma)$ (curved vertical lines). Tangent 
directions to these orbits are called `vertical', the 
$\mathcal{G}$-orthogonal directions `horizontal'. Horizontal and 
vertical directions intersect whenever the `hyper-light-cone'
touches the vertical directions, as in point $h'$. At $h,h'$, and $h''$
the vertical direction is depicted as time-, light-, and spacelike
respectively. Hence $[h']$ corresponds to a transition point where 
the signature of the metric in superspace changes.     
\label{fig:Nico5}}
\end{figure}
The diffeomorphism constraint (\ref{eq:DiffConstraint3}) for 
$j^a=0$ -- a case to which we now restrict for simplicity -- now 
simply says that $V$ must be $\mathcal{G}$--orthogonal to such 
vertical directions. We call such orthogonal directions 
\emph{horizontal}.  Moreover, it is easily seen that the inner product 
(\ref{eq:DeWittMetric}) is invariant under $\text{Diff}(\Sigma)$.
All this suggests how to endow superspace, $\mathcal{S}(\Sigma)$,
with a natural metric: take two tangent vectors at a point $[h]$
in $\mathcal{S}(\Sigma)$, lift them to horizontal vectors at 
$h$ in $\text{Riem}(\Sigma)$ and there take the inner product 
according to (\ref{eq:DeWittMetric}). 

However, this procedure only works 
if the horizontal subspace of $T_h(\text{Riem}(\Sigma))$ is truly 
complementary to the vertical space of gauge directions. However, 
this is not guaranteed due to $\mathcal{G }$ not being positive 
definite: whenever there are vertical directions of zero 
$\mathcal{G}$--norm, there will be non-trivial intersections 
of horizontal and vertical spaces. Sufficient conditions on $h$ 
for this \emph{not} to happen can be derived, like, for example, a strictly 
negative Ricci tensor~\cite{Giulini:1995b}. The emerging picture is 
that there are open sets in $\mathcal{S}(\Sigma)$ in which well 
defined hyper-Lorentzian geometries exist, which are separated by 
closed transition regions in which the signature of these metrics 
change. The transition regions precisely consist of those geometries 
$[h]$ which possess vertical directions of zero $\mathcal{G}$--norm;
see Fig.\,\ref{fig:Nico5}.

\section{Quantum geometrodynamics\index{quantum geometrodynamics}}

Einstein's theory of GR has now been brought into a form where it can be 
subject to the procedure of canonical quantization. As we have argued above, 
all the information that is needed is encoded in the constraints 
\eqref{eq:HamConstraint3} and \eqref{eq:DiffConstraint3}. However, 
quantizing them is far from trivial~\cite{Kiefer:2004}. One might 
first attempt to solve the constraints on the classical level and then 
to quantize only the reduced, physical, degrees of freedom. This is 
already impossible in quantum electrodynamics (except the case of freely 
propagating fields), and it is illusory to achieve in GR. One therefore 
usually follows the procedure proposed by Dirac and tries to implement 
the constraints as conditions on physically allowed wave functionals. 
The constraints \eqref{eq:HamConstraint3} and \eqref{eq:DiffConstraint3} 
the become the quantum conditions
\begin{eqnarray}
\label{quantumconstraints}
\hat{H}\Psi=0\ , \\
\hat{D}^a\Psi=0 \ ,
\end{eqnarray}
where the `hat' is a symbolic indication for the fact that the
classical expressions have been turned into operators. 
This procedure also applies if other variables instead of the three-metric
and its momentum are used; for example, such quantum constraints 
also play the role in loop quantum gravity, cf. the contributions of
Nicolai and Peeters as well as Thiemann to this volume. 
In the present case the resulting formalism is called quantum geometrodynamics.

Quantum geometrodynamics is defined by the transformation of
$h_{ab}(x)$ into a multiplication operator and
$\pi^{cd}$ into a functional derivative operator,
$\pi^{cd}\to -{\rm i}\hbar\delta/\delta h_{cd}(x)$.
The constraints \eqref{eq:HamConstraint3} and \eqref{eq:DiffConstraint3}
then assume the form, restricting here to the vacuum case for
simplicity,
\begin{eqnarray}
\hat{H}\Psi&\equiv&
\left(-2\kappa\hbar^2G_{abcd}\frac{\delta^2}{\delta h_{ab}\delta h_{cd}}
-(2\kappa)^{-1}\,\sqrt{h}\bigl(\,{}^{(3)}\!R-2\Lambda\bigr)\right)\Psi=0\ , 
\label{wdw}\\
\hat{D}^a\Psi &\equiv& -2\nabla_b\frac{\hbar}{\rm i}
\frac{\delta\Psi}{\delta h_{ab}} =0\ . \label{momentum}
\end{eqnarray}
Equation \eqref{wdw} is called the {\em Wheeler--DeWitt equation}
\index{Wheeler--DeWitt equation}
in honour of the work by Bryce DeWitt and John Wheeler; see e.g. 
\cite{Kiefer:2004} for details and references.
In fact, these are again infinitely many equations (one equation per
space point). The constraints \eqref{momentum}
are called the {\em quantum diffeomorphism (or momentum) constraints}. 
Occasionally, both (\ref{wdw}) and (\ref{momentum}) are referred to as
Wheeler--DeWitt equations. In the presence of non-gravitational fields,
these equations are augmented by the corresponding terms.

The argument of the wave functional $\Psi$ is the three-metric
$h_{ab}(x)$ (plus non-gravitational fields). However, because of
\eqref{momentum}, $\Psi$ is invariant under coordinate transformations
on three dimensional space (it may acquire a phase with respect
to `large diffeomorphisms' that are not connected with the
identity). A most remarkable feature of the quantum constraint equations
is their `timeless' nature -- the external parameter $t$ has 
completely disappeared.\footnote{In the case of asymptotic spaces such
a parameter may be present in connection with Poincar\'e transformations
at spatial infinity. We do not consider this case here.}
Instead of an external time one may consider an `intrinsic time'
that is distinguished by the kinetic term of \eqref{wdw}.
As can be recognized from the signature of the DeWitt metric 
\eqref{eq:WdWmetric1}, the Wheeler--DeWitt equation is
locally hyperbolic, that is, it assumes the form of a local wave equation.
The intrinsic timelike direction is related to the conformal part of the
three-metric.
With respect to the discussion in the last section
one may ask whether there are regions in superspace 
where the Wheeler--DeWitt metric exists and has precisely one negative 
direction. In that case the Wheeler--DeWitt equation would be 
strictly hyperbolic (rather than ultrahyperbolic) in a neighbourhood 
of that point. It has been shown that such regions indeed exist and 
that they include neighbourhoods of the standard round three-sphere 
geometry~\cite{Giulini:1995b}. This implies that the full Wheeler--DeWitt 
equation that describes fluctuations around the positive curvature 
Friedmann universe is strictly hyperbolic. In this case the scale factor
of the Friedmann universe could serve as an intrinsic time.
The indefinite nature of the kinetic term reflects the fact that
gravity is attractive~\cite{Giulini.Kiefer:1994}.  

There are many problems associated with the quantum constraints
\eqref{wdw} and \eqref{momentum}. 
An obvious problem is the `factor-ordering problem':
the precise form of the kinetic term
is open -- there could be additional terms proportional to $\hbar$
containing at most first derivatives in the metric. Since second functional
derivatives at the same space point usually lead to undefined
expressions such as $\delta(0)$, a regularization (and perhaps
renormalization) scheme has to be employed. Connected with this is
the potential presence of anomalies, cf. the contribution 
by Nicolai and Peeters.  
Another central problem is what choice of Hilbert space one has to
make, if any, for the interpretation of the wave functionals. 
No final answer to this problem is available in this 
approach~\cite{Kiefer:2004}.

What about the semiclassical approximation \index{emiclassical approximation}
and the recovery of an appropriate external time parameter in some limit?
For the full quantum constraints this can at least be achieved
in a formal sense (i.e., treating functional derivatives as if they were 
ordinary derivatives and neglecting the problem of anomalies); 
see~\cite{Kiefer:2004,Kiefer:2006}. The discussion is also connected to the 
question: Where does the imaginary unit i in the 
(functional) Schr\"odinger equation come from?
The full Wheeler--DeWitt equation is real,
and one would thus also expect real solutions for $\Psi$.
An approximate solution is found through a Born--Oppenheimer-type of
scheme, in analogy to molecular physics. The state then assumes the form
\begin{equation}
\label{expiS}
\Psi \approx \exp({\rm i}S_0[h]/\hbar) \, \psi[h,\phi]\ ,
\end{equation}
where $h$ is an abbreviation for the three-metric, and
$\phi$ stands for non-gravitational fields. In short, one finds that
\begin{itemize}
\item $S_0$ obeys the Hamilton--Jacobi equation for the gravitational field
and thereby defines a classical spacetime which is a solution to
Einstein's equations (this order is formally similar to the recovery
of geometrical optics from wave optics via the eikonal equation).
\item $\psi$ obeys an approximate (functional) Schr\"odinger equation,
\begin{equation}
\label{semi}
 {\rm i}\hbar \, \underbrace{\nabla \, S_0 \, \nabla
\psi}_{\frac{\displaystyle\partial\psi}{\displaystyle\partial t}} 
\approx H_{\rm m} \, \psi \ ,
\end{equation}
where $H_{\rm m}$ denotes the Hamiltonian for the non-gravitational fields
$\phi$. Note that the expression on the left-hand side of \eqref{semi} 
is a shorthand notation for an integral over space, in which $\nabla$
stands for functional derivatives with respect to the three-metric. 
Semiclassical time $t$ is thus defined in this limit from the
dynamical variables. 
\item The next order of the Born-Oppenheimer scheme yields
 quantum gravitational correction terms proportional to the inverse
Planck mass squared, ${1}/{m_{\rm P}^2}$. The presence of such terms
may in principle lead to observable effects, for example, in the
anisotropy spectrum of the cosmic microwave background radiation. 
\end{itemize}

The Born--Oppenheimer expansion scheme distinguishes a state of the form
\eqref{expiS} from its complex conjugate. In fact, in a generic situation
both states will decohere from each other, that is, 
they will become dynamically independent~\cite{Joos.etal:2003}. 
This is a type of symmetry breaking, in analogy to the
occurrence of parity violating states in chiral molecules. It is through
this mechanism that the i and the $t$
in the Schr\"odinger equation emerge.

The recovery of the Schr\"odinger equation \eqref{semi} raises an
interesting issue. It is well known that the notion of Hilbert space
is connected with the conservation of probability (unitarity) and thus
with the presence of an external time (with respect to which the
probability is conserved). The question then arises whether the concept
of a Hilbert space is still required in the {\em full} theory where
no external time is present. It could be that this concept makes sense
only on the semiclassical level where \eqref{semi} holds.

\section{Applications}

The major physical applications of quantum gravity concern
cosmology and black holes. Although the above presented formalism
exists, as yet, only on a formal level, one can study models
that present no mathematical obstacles. Typically, such models are
obtained by imposing symmetries on the equations~\cite{Kiefer:2004}.
Examples are spherical symmetry (useful for black holes) and
homogeneity and isotropy (useful for cosmology). 

Quantum cosmology is the application of quantum theory to the
universe as a whole. 
Let us consider a simple example: a
Friedmann universe with scale factor $a\equiv
  {\rm e}^{\alpha}$ containing a
massive scalar field {$\phi$}. In this case,
the diffeomorphism constraints \eqref{momentum}
are identically fulfilled, and the
Wheeler--DeWitt equation \eqref{wdw} reads 
\begin{equation}
\label{mini}
\hat{H}\psi\equiv\left({G}\hbar^2\frac{\partial^2}
{\partial \alpha^2}
-\hbar^2\frac{\partial^2}{\partial\phi^2}
+m^2\phi^2{\rm e}^{6\alpha}-\frac{{\rm e}^{4\alpha}}{{G}}
\right)\psi(\alpha,\phi)=0\ .
\end{equation}
This equation is simple enough to find solutions (at least numerically)
and to study physical aspects such as the dynamics of wave packets
and the semiclassical limit~\cite{Kiefer:2004}.

There is
one interesting aspect in quantum cosmology that possesses
far-reaching physical consequences.
Because \eqref{wdw} does not contain an
external time parameter $t$, the quantum theory
exhibits a kind of determinism drastically different from the 
classical theory~\cite{Zeh:2001}\cite{Kiefer:2004}.
Consider a model with a two-dimensional
configuration space spanned by the scale factor, $a$, and a homogeneous
scalar field, $\phi$, see Fig.~6.
 (Such a model is described,
for example, by \eqref{mini} with $m=0$.) 
The classical
model be such that there are solutions where the universe expands
from an initial singularity, reaches a maximum, and recollapses to a
final singularity. Classically, one would impose,
in a Lagrangian formulation, $a,\dot{a},\phi,
\dot{\phi}$ (satisfying the constraint) at some $t_0$ (for example, 
at the left leg of the trajectory), and then the trajectory would be
determined. This is indicated on the left-hand side of 
Fig.~6. 
\begin{figure}[h]
\label{fig:fig_berlin05_6ab}
\caption{The classical and the quantum theory of gravity exhibit
drastically different notions of determinism.}
\begin{center}
\includegraphics[width=5cm]{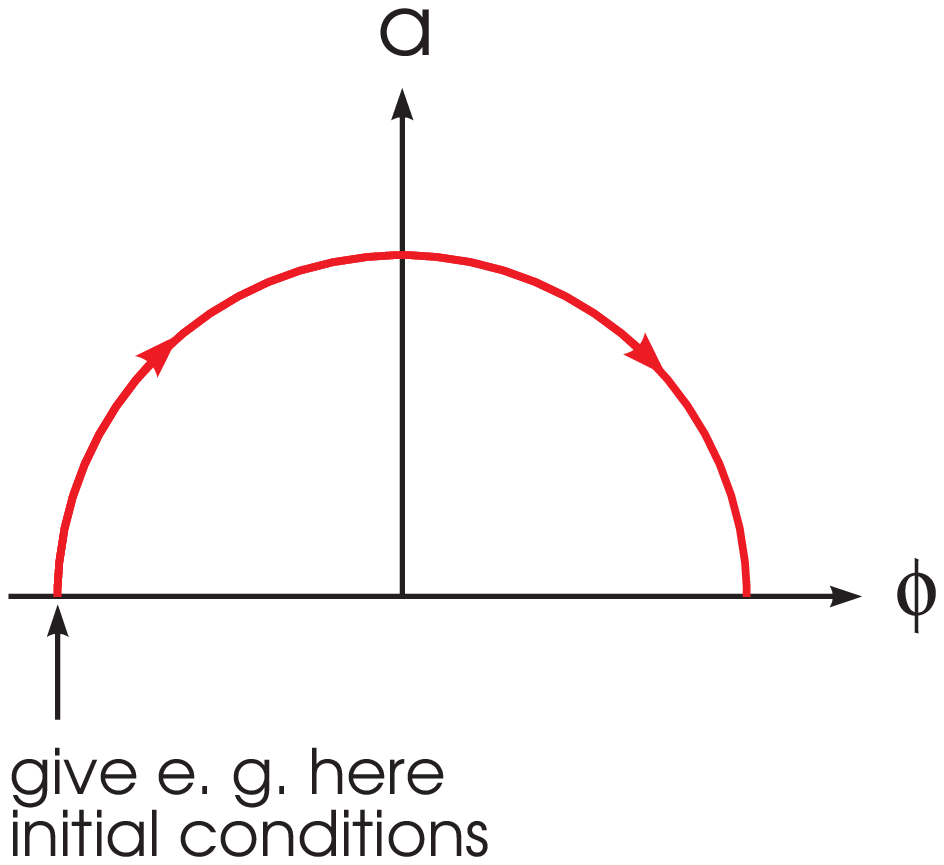}
\includegraphics[width=5cm]{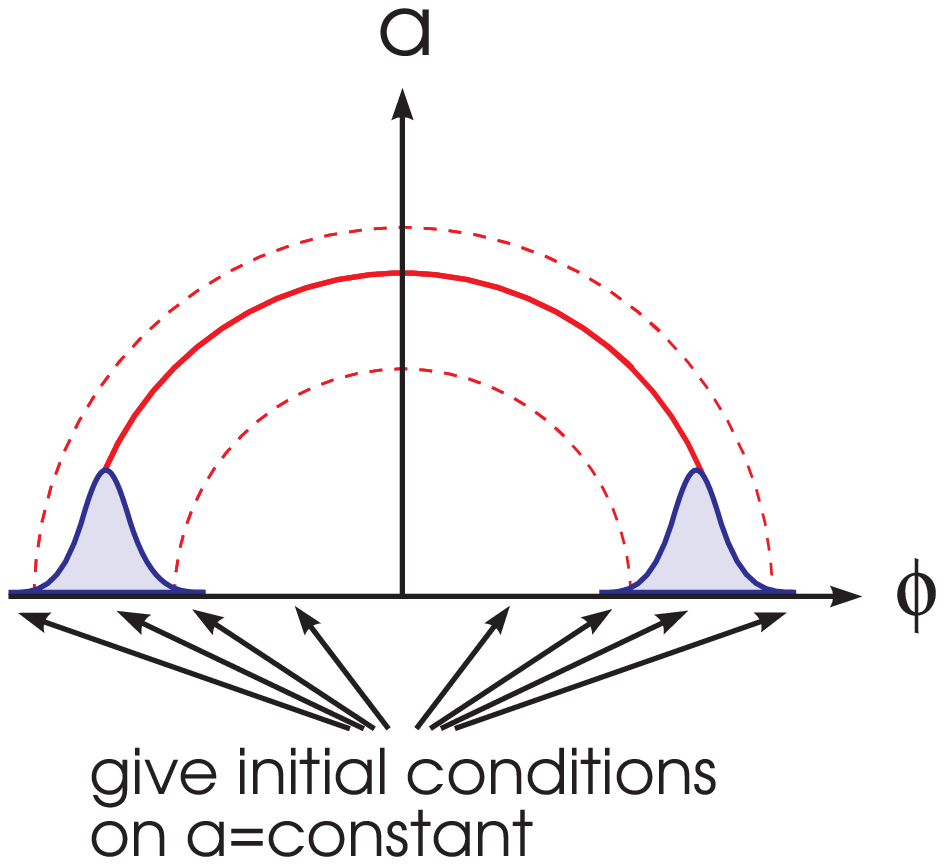}
\end{center}
\end{figure}
In the quantum theory, on the other hand, there is no $t$.
The hyperbolic nature of a minisuperspace equation such as
\eqref{mini} suggests to impose boundary conditions
at $a=$ constant. In order to represent the classical trajectory
by narrow wave packets, the `returning part' of the packet must
be present `initially' (with respect to $a$). The determinism of the
quantum theory then proceeds from small $a$ to large $a$, not along
a classical trajectory (which does not exist). This behaviour has
consequences for the validity of the semiclassical approximation
and the arrow of time. In fact, it may in principle be possible
to understand the origin of irreversibility from quantum cosmology,
by the very fact that the Wheeler--DeWitt equation is
asymmetric with respect to the intrinsic time 
given by $a$. The framework of canonical quantum cosmology
is also suitable to address the quantum-to-classical transition
for cosmological variables such as the volume of the 
universe~\cite{Joos.etal:2003}\cite{Kiefer:2004}.
Using the approach of loop quantum gravity (see Thiemann's contribution)
one arrives at a Wheeler--DeWitt equation in cosmology which is
fundamentally a difference equation instead of a differential equation
of the type \eqref{mini}. In the ensuing framework of loop
quantum gravity it seems that the classical singularities of GR
can be avoided.

Singularity avoidance for collapse situations can also be found
from spherically symmetric models of quantum geometrodynamics.
For example, in a model with a collapsing null dust cloud,
an initially collapsing wave packet evolves into a superposition
of collapsing and expanding packet~\cite{Hajicek:2003}.
This leads to destructive interference at the place where the
singularity in the classical theory occurs. 
Other issues, such as the attempt to give a microscopic derivation
of the Bekenstein--Hawking entropy (see the contribution by C.~Kiefer
to this volume), have been mainly addressed in loop quantum gravity.
A final, clear-cut, derivation remains, however, elusive.


\end{document}